# Testing the accuracy of Coimbra Astronomical Observatory solar filament historical series (1929-1941)

Ana Lourenço, Ricardo Gafeira, Vitor Bonifácio, Teresa Barata, João Fernandes, Eva Silva

Ana Lourenço: Univ Coimbra, Centre for Earth and Space Research of the University of Coimbra, Department of Earth Sciences; OGAUC–Geophysical and Astronomical Observatory of the University of Coimbra, Rua do Observatório, s/n, 3040–004 Coimbra, Portugal; ana.lourenco@dct.uc.pt
Tel.: +351 239 802 272

Ricardo Gafeira: Univ Coimbra, Instituto de Astrofísica e Ciências do Espaço; OGAUC–Geophysical and Astronomical Observatory of the University of Coimbra, Rua do Observatório s/n, 3040-004 Coimbra, Portugal
Instituto de Astrofísica de Andalucía (CSIC), Apartado de Correos 3004, E-18080 Granada, Spain

Vitor Bonifácio: Physics Department, University of Aveiro, Research Centre "Didactics and Technology in Education of Trainers" (CIDTFF), Campus Universitário de Santiago, 3810-193 Aveiro, Portugal

Teresa Barata: Univ Coimbra, Instituto de Astrofísica e Ciências do Espaço; OGAUC–Geophysical and Astronomical Observatory of the University of Coimbra, Rua do Observatório, s/n, 3040–004 Coimbra, Portugal

João Fernandes: Univ Coimbra, Centre for Earth and Space Research of the University of Coimbra, Rua do Observatório, s/n, 3040–004 Coimbra, Portugal; Department of Mathematics, University of Coimbra, 3000-456 Coimbra, Portugal

Eva Silva: Univ. Coimbra, Department of Mathematics, 3000-456 Coimbra, Portugal

**Abstract**

*The present work aims to validate the positions of solar filaments published in the Annals of Coimbra University Astronomical Observatory, currently the Geophysical and Astronomical Observatory of the University of Coimbra, corresponding to years 1929 to 1941. The published Stonyhurst positions were obtained by an original method devised in the early 20th century that used a spherical calculator instrument, a wood-made model of the sun. We used the digital images of the original spectroheliograms to measure the positions of the filaments, and heliographic coordinates were determinate with the routines implemented on python package Sunpy. The correlation coefficients between both sets of coordinates are positive and highly significant. The results validate the method used at the Coimbra observatory and the data published. We conclude that Coimbra solar filaments catalogues are reliable and can therefore be considered for future solar activity studies.*

**Keywords** solar filaments, Stonyhurst heliographic coordinates, historical data, catalogues.

## 1. Introduction

Long-term records are essential to understand solar activity since its processes may take decades or centuries to reveal themselves. Therefore, several observatories around the world started continuous solar observation programs (*e.g.*, Mouradian and Garcia, 2007). Known examples of long series are those of the Royal Greenwich Observatory, UK (Willis et al., 2013), Meudon, France (Mein and Ribes, 1990), Kodaikanal, India (Mandal et al., 2017), Mount Wilson, USA (Lefebvre et al., 2005), Debrecen, Hungary (Baranyi et al., 2016) and Arcetri, Italy (Ermolli et al., 2009). Other programs were also implemented in Ebro and San Fernando in Spain (Aparicio et al., 2014; Curto et al., 2016) and Coimbra, Portugal (Mouradian and Garcia, 2007; Garcia et al. 2010a and 2010b, 2011; Carrasco et al., 2018a and 2018b). The Coimbra Astronomical Observatory (currently Geophysical and Astronomical Observatory of the University of Coimbra, "Observatório Geofísico e Astronómico da Universidade de Coimbra", hereafter OGAUC) started in 1925 a solar observation program that continues today, with essentially the same acquisition system leading to a large and consistent database. That program led to the publication in 1932 of the first volume of the Observatory Annals analysing the 1929 solar activity. An immense endeavour that was well received by the international community at the time (Deslandres, 1932; Bonifácio, 2017). The overstretched observatory staff managed to continue to publish volumes analysing the years 1929 to 1944 regularly, at first, and with longer and longer intervals after the Second World War. Finally, a volume for 1979 was published in 1986. In this paper, we aim to validate the solar filament data published. This work is part of a broader project that aims to recover and validate OGAUC historical information on solar observations. Recently, several solar databases from observatories around the world, including Coimbra (Carrasco et al., 2018b), Ebro (Curto et al., 2016), Greenwich (Willis et al., 2013), Locarno (Cortesi et al., 2016), Madrid (Lefèvre et al. 2016), Kodaikanal (Mandal et al., 2017) and Valencia (Carrasco et al., 2014) have been digitized and analysed. However, most solar data acquired before the second half of the 20th century has not been digitized which makes it unavailable to the scientific community (Lefèvre and Clette, 2014). The urgent need for historical astronomical data digitization and preservation enabling scientific research is referred by Pevtsov et al. (2019). Carrasco et al. (2018a) provided a machine-readable version of sunspots data recorded in the Coimbra historical solar catalogues. Data on other solar features have not yet been addressed. The work of recovering historical solar observations needs to be continued to complete temporal coverage of the observations and to correct possible gaps in the data series (Lefèvre and Clette, 2014). The present work intends to participate in this collective effort to recover historical solar databases.

Firstly, we analysed the method used at the OGAUC to obtain the filament solar coordinates from the original spectroheliograph CaII K3 photographs. This led us to study the advantages and limitations of, to the best of our knowledge, a kind of solar heliographic calculator, commonly known as Costa Lobo's sphere, invented by the then director Francisco Costa Lobo (1864-1945). Secondly, we selected a set of CaII K3 spectroheliograms from the years 1929 to 1941 and read the corresponding filament coordinates in the catalogues. Finally, these were compared with the coordinates obtained via the automatic method described in Duffett-Smith and Zwart (2011). Solar filaments were chosen instead of sunspots because it is more accurate to locate points, such as the filaments footpoints. These solar structures appear from equator to pole during all phases of the solar cycle and are potential representatives of solar magnetic activity (McIntosh, 1972), therefore they are used to locate the positions of neutral magnetic lines on the synoptic maps (Peng, 2013). In this paper, we concentrate on the determination of the location of filament endpoints/footpoints. It is currently accepted that footpoints of a filament and the magnetic field line "are co-spatial on the same magnetic polarity" (Chen et al., 2020). Moreover, it is known how relevant endpoints locations are in order to determine the real filament spine, particularly used on the automated detection methods (Hao et al., 2018). On the other hand, those points are important to filament chirality studies (Hao et al., 2016). In addition to the solar physics context study, this type of structure is important for Space Weather studies. Solar filaments play a vital role since they can indicate the possible occurrence of CMEs (Atoum, 2016), which are the major driving source of the hazardous space weather around the Earth (Chen, 2011).

## 2. Solar observations at the OGAUC

The daily observations at the OGAUC started in December 1925 and extend till today approximately 9 solar cycles. Images on CaII K1 and CaII K3 spectral lines are obtained daily, depending on the weather condition and the publication of these images started in January 1926 (Lobo, 1932). The solar images are acquired since the beginning with the same instrument, a spectroheliograph equivalent to the one installed at Meudon Observatory in Paris, which has been subject to improvements over the years (Bualé et al., 2007; Klvaňa et al., 2007; Garcia et al., 2010a). The spectroheliograph has a focal length of 4060 mm and a diameter of 200 mm, which produces a maximum diameter of the solar disk on the entrance slit of 38.5 mm. After the entrance still, the optical system consists of a collimator with a focal length of 1300 mm and 150 mm diameter. At the time of the analysed observations, the dispersion device was a set three glass prism that allowed a spectral resolution of 0.25 nm at 3933 nm. Although the telescope has a theoretical diffraction-limited spatial resolution under 1", when considering the slit width and average atmospheric conditions, the estimated resolution of the system is between 1.5" and 2".

Until the present day, more than 50 000 spectroheliograms were acquired by the OGAUC spectroheliograph (Lourenço et al., 2019). Recently, software tools have been developed to automatically detect and analyse chromospheric plages (Barata et al., 2018) and sunspots (Carvalho et al., 2020) on solar images acquired at the OGAUC.

The OGAUC catalogue uses Stonyhurst heliographic system solar coordinates. The complete procedure started by drawing the solar features directly from a spectroheliogram (Figure 1) into a superposed normal azimuthal projection grid centred at the image centre (Figure 2).

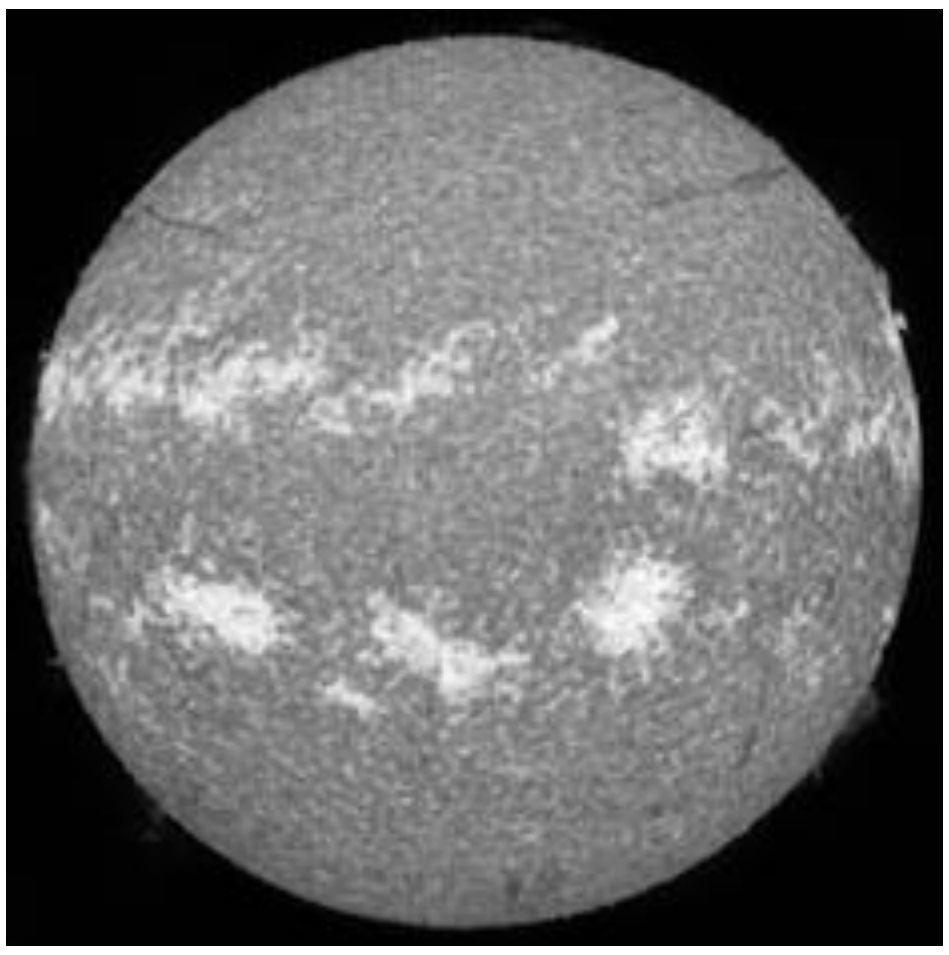

**Figure 1** Ca K3 spetroheliogram: 13 September 1938. Plages and filaments can be seen on the solar disk.

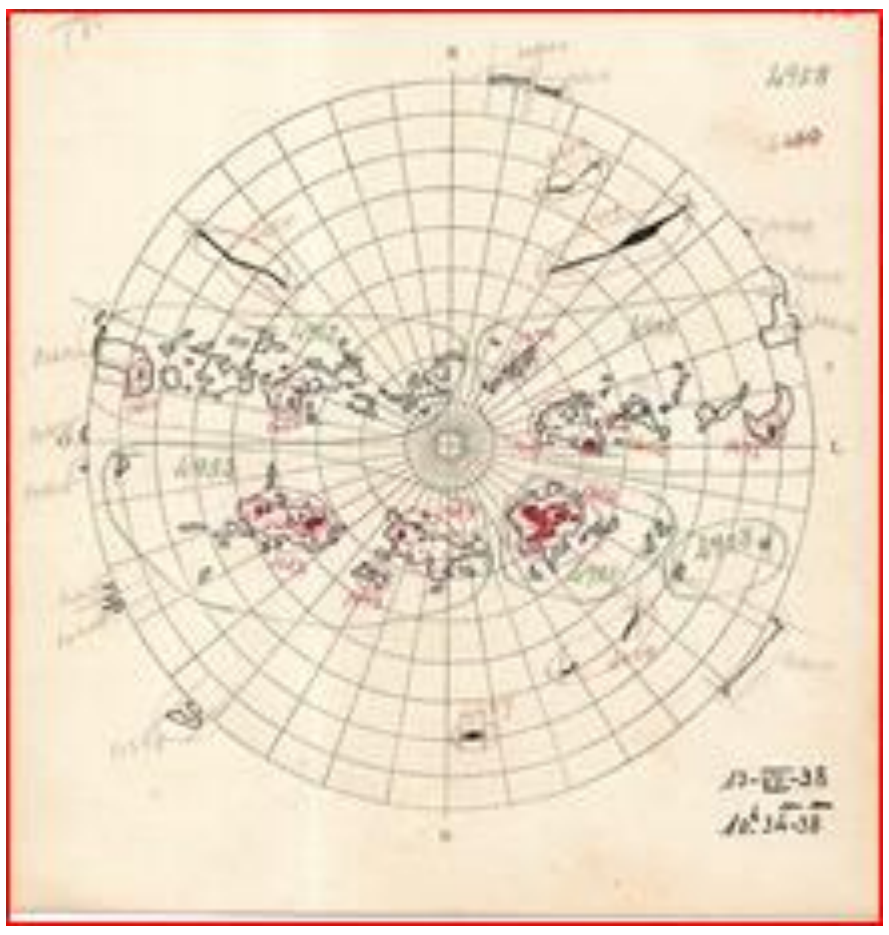

**Figure 2** Plages and filaments drawings below an azimuth grid, correspondent to Figure 1.

The solar Stonyhurst longitude, P, and angles CA were read in the grid (Figure 3). Knowing the heliographic latitude of the central point of the solar disk (B0) by the ephemerides solar pole may be marked in the grid so arc CPolo (=90-B0) is also known (Figure 3). So, using spherical trigonometry we can determine arc PoloA, that is the heliographic co-latitude of A, from the spherical triangle with vertices {A, Polo, C}. However, this procedure involves the computation of ten circular functions (sine or cosine) per point. At those times, these computations were particularly laborious (using interpolation in tables with several decimal places). On the other hand, for each spectroheliogram, one could have 20 or more interesting points.

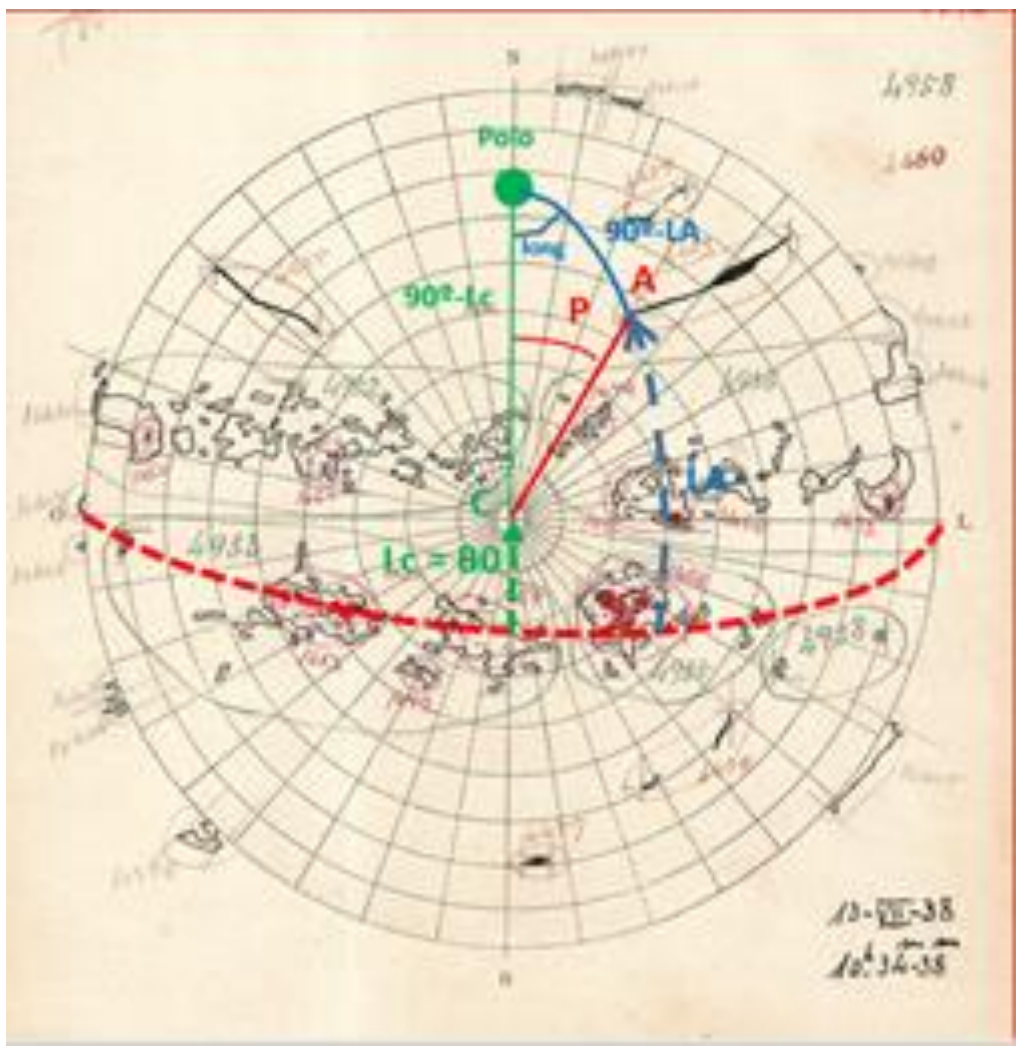

**Figure 3** Ibidem Figure 2 to extract the angle P and the major arc circle CA used for the spherical trigonometry calculations. Dashed red line - equator; dashed blue line - heliographic latitude; Lc is the image centre latitude.

Francisco Costa Lobo developed a novel method to facilitate and speed up the calculation of the coordinates (Lobo, 1932; http://193.137.102.29/ObservatorioAstronomicoMuseu/entity_detail.aspx?aid=1135). He designed a device, named "Esfera Solar de Costa Lobo" (Costa Lobo's Solar Sphere), a wooden sphere with a graduated metallic armature superimposed (Figure 4). The sphere's diameter of 573 mm was chosen due to the dimensions of the OGAUC spectroheliograph images. As such 0.5 millimetre of the maximum circle corresponds to 0.1 degrees, i.e., 6'. The sphere then allows the transformation of the position angle and distance coordinates to solar coordinates with a precision of 0.1º. Costa Lobo's sphere had a metallic armature that defined a spherical triangle, in figure 4B the triangle with sides *PA*, *AB* and *BP*. The arms *PB* and *AB* are movable, using hinges, exactly around points *P* and *A* positioned at an angular distance of 90° from each other. One starts by moving arm *PA* along the *PP1* arc. Point *P* can then be located at the position of the Sun pole on the day of the observation, so that point

*A* corresponds exactly to the Sun centre. Then, moving arm *AB* around point *A*, the technician point *A* on the sphere, with a stiletto. By moving *PB* around *P*, until it touches the signal previously marked, heliographic co-latitude and longitude are read on arms *PB* and *AB,* respectively.

One should point out that the sphere did not waive altogether the use of appropriate tables, for instance for points overlaying the sphere metallic arms but it greatly reduced the time and number of employees needed (Lobo, 1934).

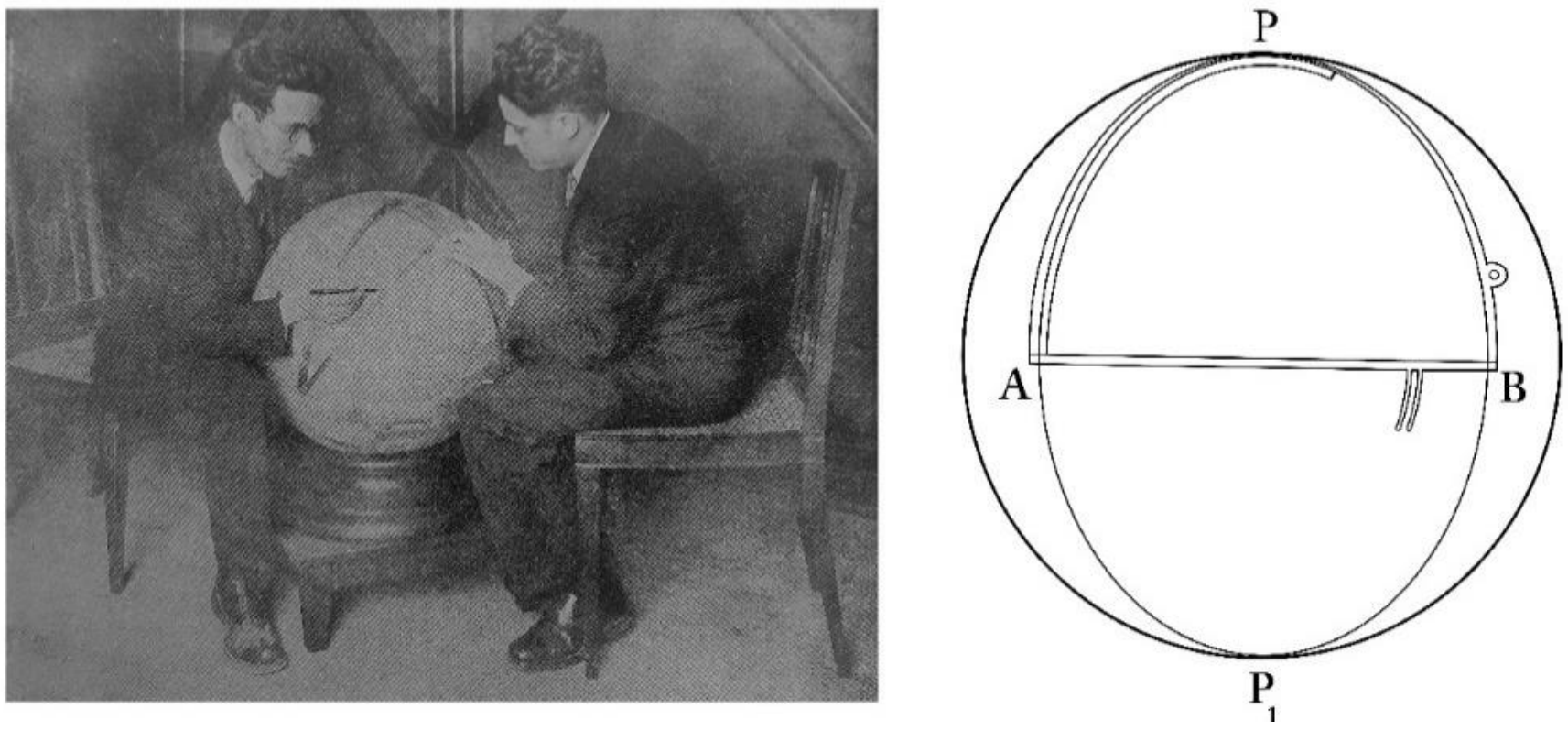

**A** **B**

**Figure 4** The Costa Lobo's Solar Sphere (A) and outline draw of the metallic armature (B) [Source: Lobo, 1934].

Note that Costa Lobo's sphere is not a universal spherical triangle calculator, as other apparatus invented at the beginning of the XX century (Morata, 2012), and should be used only in the context described in this work. The main reason is that the arc C Polo length is constrained to 90.00 ± 7.25. On the other hand, in a few situations when its use is not possible standard spherical trigonometry computations are then used. In Figure 5 one can perceive the method used as well as the amount of work required.

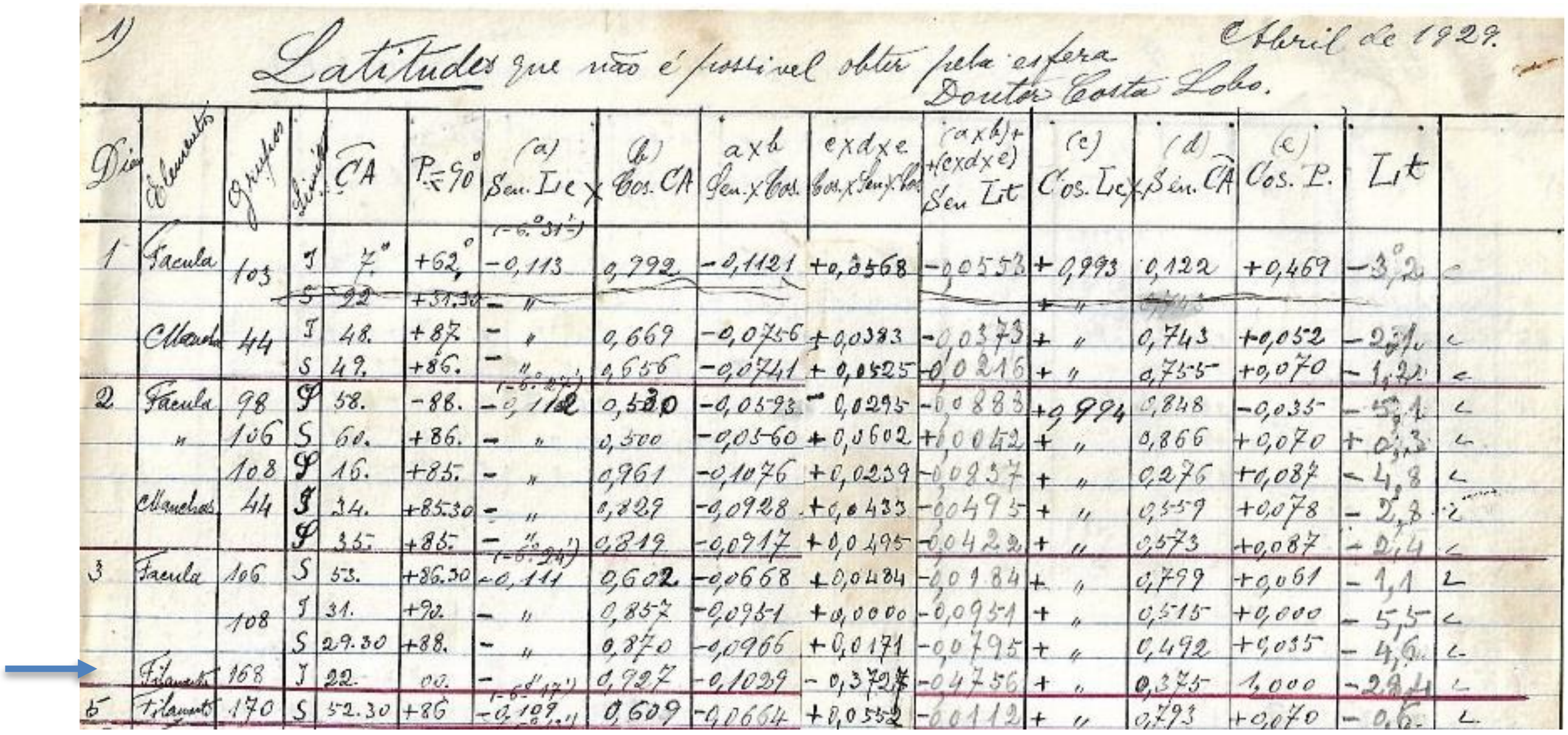

1) Latitudes que não é possível obter pela esfera Doutor Costa Lobo. Abril de 1929.

| Dia | Elementos | Grupos | Limite | CA | P∓90° | (a) Sen. Lc | (b) Cos. CA | a×b Sen.×Cos. | c×d×e Cos.×Sen.×Cos. | (a×b)+(c×d×e) Sen Lt | (c) Cos. Lc | (d) Sen. CA | (e) Cos. P. | Lt |
|---|---|---|---|---|---|---|---|---|---|---|---|---|---|---|
| 1 | Facula | 103 | I | 7° | +62° | (-6°31') -0,113 | 0,792 | -0,1121 | +0,3568 | -0,0553 | +0,993 | 0,122 | +0,469 | -3,2 |
| | | | S | 22 | +51,30 | - " | | | | | + " | | | |
| | Manchas | 44 | I | 48. | +87 | - " | 0,669 | -0,0756 | +0,0383 | -0,0373 | + " | 0,743 | +0,052 | -2,1 |
| | | | S | 49. | +86. | - " | 0,656 | -0,0741 | +0,0525 | -0,0216 | + " | 0,755 | +0,070 | -1,22 |
| 2 | Facula | 98 | I | 58. | -88. | (-6°24') -0,112 | 0,530 | -0,0593 | -0,0295 | -0,0883 | +0,994 | 0,848 | -0,035 | -5,1 |
| | " | 106 | S | 60. | +86. | - " | 0,500 | -0,0560 | +0,0602 | +0,0042 | + " | 0,866 | +0,070 | +0,3 |
| | | 108 | I | 16. | +85. | - " | 0,961 | -0,1076 | +0,0239 | -0,0837 | + " | 0,276 | +0,087 | -4,8 |
| | Manchas | 44 | I | 34. | +85.30 | - " | 0,829 | -0,0928 | +0,0433 | -0,0495 | + " | 0,559 | +0,078 | -2,8 |
| | | | I | 35. | +85. | - " | 0,819 | -0,0917 | +0,0495 | -0,0422 | + " | 0,573 | +0,087 | -2,4 |
| 3 | Facula | 106 | S | 53. | +86.30 | (-6°24') -0,111 | 0,602 | -0,0668 | +0,0484 | -0,0184 | + " | 0,799 | +0,061 | -1,1 |
| | | 108 | I | 31. | +90. | - " | 0,857 | -0,0951 | +0,0000 | -0,0951 | + " | 0,515 | +0,000 | -5,5 |
| | | | S | 29.30 | +88. | - " | 0,870 | -0,0966 | +0,0171 | -0,0795 | + " | 0,492 | +0,035 | -4,6 |
| | Filamento | 168 | I | 22. | 00. | - (-6°17') | 0,927 | -0,1039 | -0,3729 | -0,4756 | + " | 0,375 | 1,000 | -28,4 |
| 5 | Filamento | 170 | S | 52.30 | +86 | -0,109 | 0,609 | -0,0664 | +0,0552 | -0,0112 | + " | 0,793 | +0,070 | -0,6 |

**Figure 5** Handwritten notes on latitude calculations for a few days of the year 1929 (the blue arrow marks the row with a calculation error described in section 4).

The solar phenomena coordinates obtained by this method were published, as previously referred, in the "Anais do Observatório Astronómico da Universidade Coimbra–Fenómenos Solares" (Annals of the Astronomical Observatory of the University of Coimbra–Solar Phenomena). These catalogues began in 1932 with the results of daily solar observations, made with the spectroheliograph for the year 1929. In 1928, a first publication was made, preceding the catalogues collection. It briefly reported a few results for the years 1926 and 1927 (Lobo, 1928). Until 1986, 16 volumes were published with data for the period 1929–1944 and one for the year of 1979 (Lourenço et al., 2019). Parameters such as position, size, and number of sunspots, facular regions, filaments, and prominences were daily reported. An example page of the OGAUC catalogues containing information on solar filaments is shown in Figure 6. The columns record the following information: 1) day and hour of the observation; 2) number of the filament assigned by the OGAUC (counting started on January 1, 1929); 3) the date when the filament was observed for the first time; 4) temporal duration from the first occurrence of the phenomenon; 5) latitude limits; 6) longitude limits; 7) filament characteristics (e.g. thin, continuous, angular) and 8) filament length in hundredths of the solar radius

DA UNIVERSIDADE DE COIMBRA 255

**Dezembro** **FILAMENTOS** **1941**

| Época T. M. G. Loc. obs. | N. | D. ap. | D. v. | Lim. lat. | | Lim. long. | | Características | Ext. Cent. do r. |
|---|---|---|---|---|---|---|---|---|---|
| | | | | ° | ° | ° | ° | | |
| 1 $11^h\ 30^m$-$34^m$ | 5988 | 1-XII | .. | −14,5 | −18,0 | +82,3 | +89,4 | f, an | 13 |
| | 5989 | » | .. | +12,7 | +15,3 | +71,7 | +83,3 | f, an, en | 21 |
| | 5990 | » | 1d | −15,2 | −21,3 | +35,8 | +43,3 | f, an | 16 |
| 2 $10^h\ 6^m$-$10^m$ | 5988 | 1-XII | .. | − 7,4 | −31,1 | +50,5 | +89,0 | fl, an | 74 |
| | 5989 | » | .. | +11,8 | +14,2 | +53,9 | +61,4 | fl, an, en | 13 |

**Figure 6** Example page of the OGAUC solar catalogue of December 1941, with filaments data [source: Reis, 1965].

The image acquisition and solar phenomena study continued, despite the suspension of the publication due to economic issues. The resulting information has been preserved but dispersed in various documents.

## 3. Validation methodology

This work is mainly focused on the results obtained from the daily CaII K3 observations acquired at the OGAUC, in the early 20$^{th}$ century. Test images were chosen from the spectroheliograms published in the catalogues for the period 1929–1941, in order to have a temporal coverage as uniform as possible. The year 1940 was not included in this set, since the drawings with the azimuthal projection grid are to the best of our knowledge lost. Each catalogue issue published at least 2 to 4 spectroheliograms images per month with interesting solar phenomena. From this set, we selected 110 images covering solar cycles 16, 17 and 18. The images - negative photos - were downloaded from the OGAUC database and transformed to positives. These images (E-W) were mirrored to be compared with the drawings (W-E) (Figure 7). This procedure allowed us to correctly identify the filaments, obtain their number and read their coordinates in the catalogue's tables. A set of 168 filaments were identified, allowing the test of 336 pairs of coordinates.

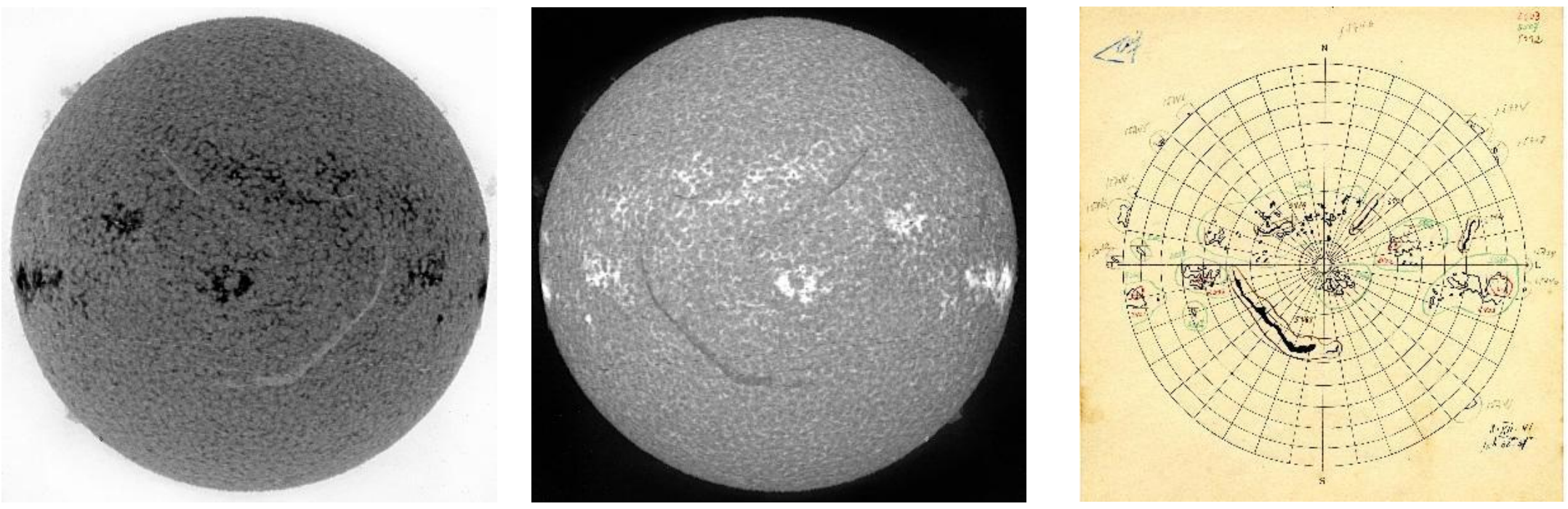

A B C

**Figure 7** Original image (A), image ready to use (B), and the correspondent azimuthal projection grid (C).

Note that in the Annals two values are recorded for each filament, representing their limits in latitude and longitude. Usually, these limits overlap the ends of the filaments (Figure 8A). In a few situations, the latitude and/or longitude limits are obtained in a middle segment of the filament (Figure 8B and 8C). We took these situations into account when using the automatic method.

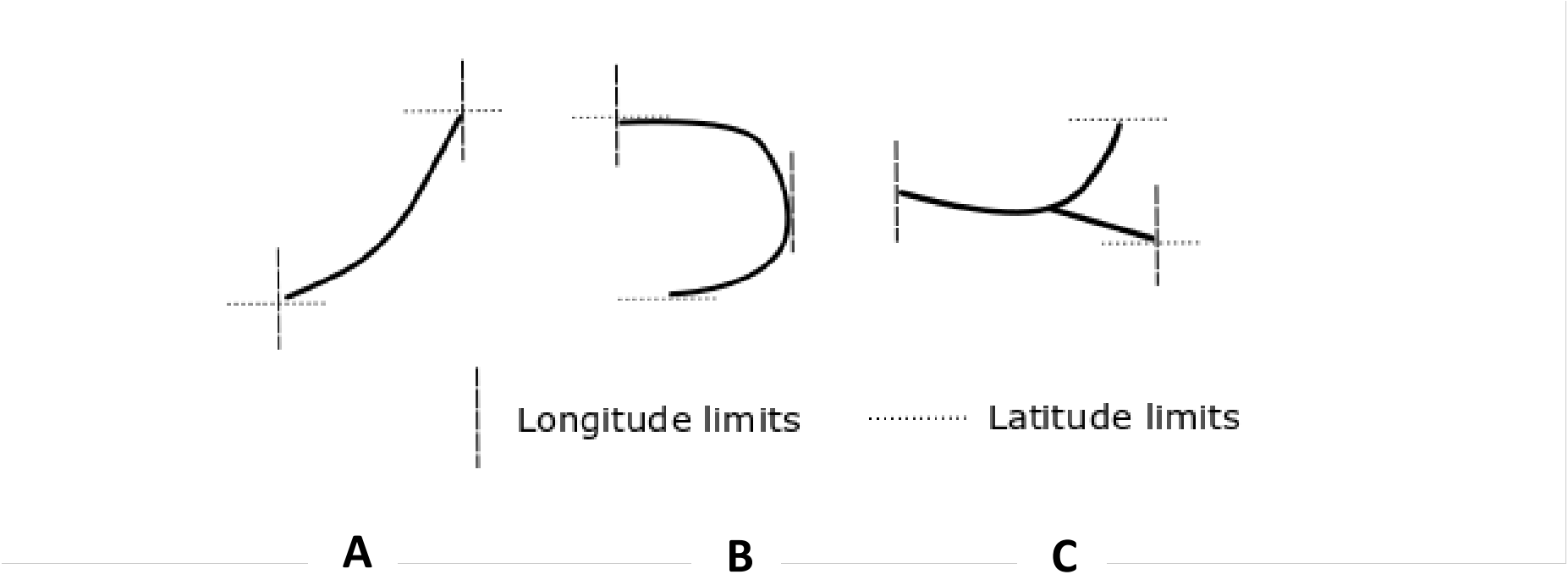

**Figure 8** Examples of filament types and processes used to mark latitude and longitude limits, on drawings of solar features.

To compute the heliographic coordinates, we started by defining the centre and diameter of the solar disk. That was done by fitting a circle to the solar disk binary image. We binarized the data with a threshold equal to two times the average intensity of a rectangle with size 30 by 30 pixels placed in a dark part of the image (bottom left corner).

Using the solar disk position and size, and the positions of the filament identified manually in the positive image (Figure 7B), we used the routines implemented on python package Sunpy (https://docs.sunpy.org) to compute the heliographic coordinates.

To estimate the error in this calculation, we assumed a miss estimation of a 2-pixel radius while manually determining the location of the points of interest in the images.

We then computed the difference between the calculated heliographic coordinates assuming the "correct" pixel and if an error of 2-pixel was committed in both direction and dimensions. We observed that the error is around 0.5° in the disk increasing to values close to 3° very close to the limb.

## 4. Results and discussion

Published and automatically calculated coordinates are listed in the appendix, as well as the images date. In order to assess the reliability of published data, the computed longitudes and latitudes were plotted versus published ones (Figure 9). A positive and highly significant correlation was found ($r$=0.99, $\rho < 0.001$) between the sets. We have identified 18 events where the longitude exceeds 90° (5% of the events, marked in bold in the Appendix table). One possible explanation for obtaining values outside the -90 to 90 interval is the fact that for filaments observed at the limb, we can still see some signal coming from the filament outside the disk, cleating like a partial prominence (Figure 10). In that case, the observers assumed an extension to the classical interval for longitude and latitude and attributed absolute values higher than 90. An approach valid in this context and used in other databases, so we decided not to exclude them. Some outliers may be due to calculation errors. For example, the published latitude of filament 168, from April 3, 1929, was wrongly calculated (figure 5, row marked with a blue arrow). A mistaken signal gives a difference of 44° between published and correctly calculated value. This also points out another advantage of using Costa Lobo's sphere, in addition to the reduction of time and number of employees needed to perform this task. Finally, one cannot rule out typographic typos. At least two obvious typographic errors were identified for the published values, one for the southern longitude of filament 3701 from April 7, 1936, and the other for the northern longitude of filament 3407 from 2 May 1936. These values are responsible for the greater differences observed between published and calculated coordinates (14.7 and 11.7, respectively, table 1). Typographic and other errors are less than 3% (marked in italic in the Appendix table). We realized that errors are mainly found in the catalogues published during and after the Second World War. Lack of specialized personnel and observatory staff mobility may be partly to blame. For instance, José António Madeira (1896-1976) chief observatory observer since 1926 left in 1942 (Madeira, 1957). Typographic, systematic, and isolated errors are common in old catalogues so recently an effort has been made to identify and correct these errors (*e.g.,* Carrasco et al. 2014; Willis et al. 2013; Erwin et al., 2013).

The basic statistic of differences between coordinates published and calculated is presented in table 1. The latitudes published differ by an average absolute value of 1.4° from those calculated automatically. Relative to the longitudes, the comparison shows higher differences, with an average absolute value of 2.1° for northern footpoints and 2.3° for southern footpoints. Except for the errors mentioned above, the maximum longitude differences were observed for points located near the solar limb, which may be due to the difficulty in determining features position near the limb (*e.g.,* Poljančić et al., 2011). Values that tie well with the previously referred error estimation.

North-south asymmetry was also analysed to verify if the sphere returns accurate results in both hemispheres. The difference between the published and automatic coordinates are slightly greater in the southern hemisphere (Table 1, columns 5 to 8).

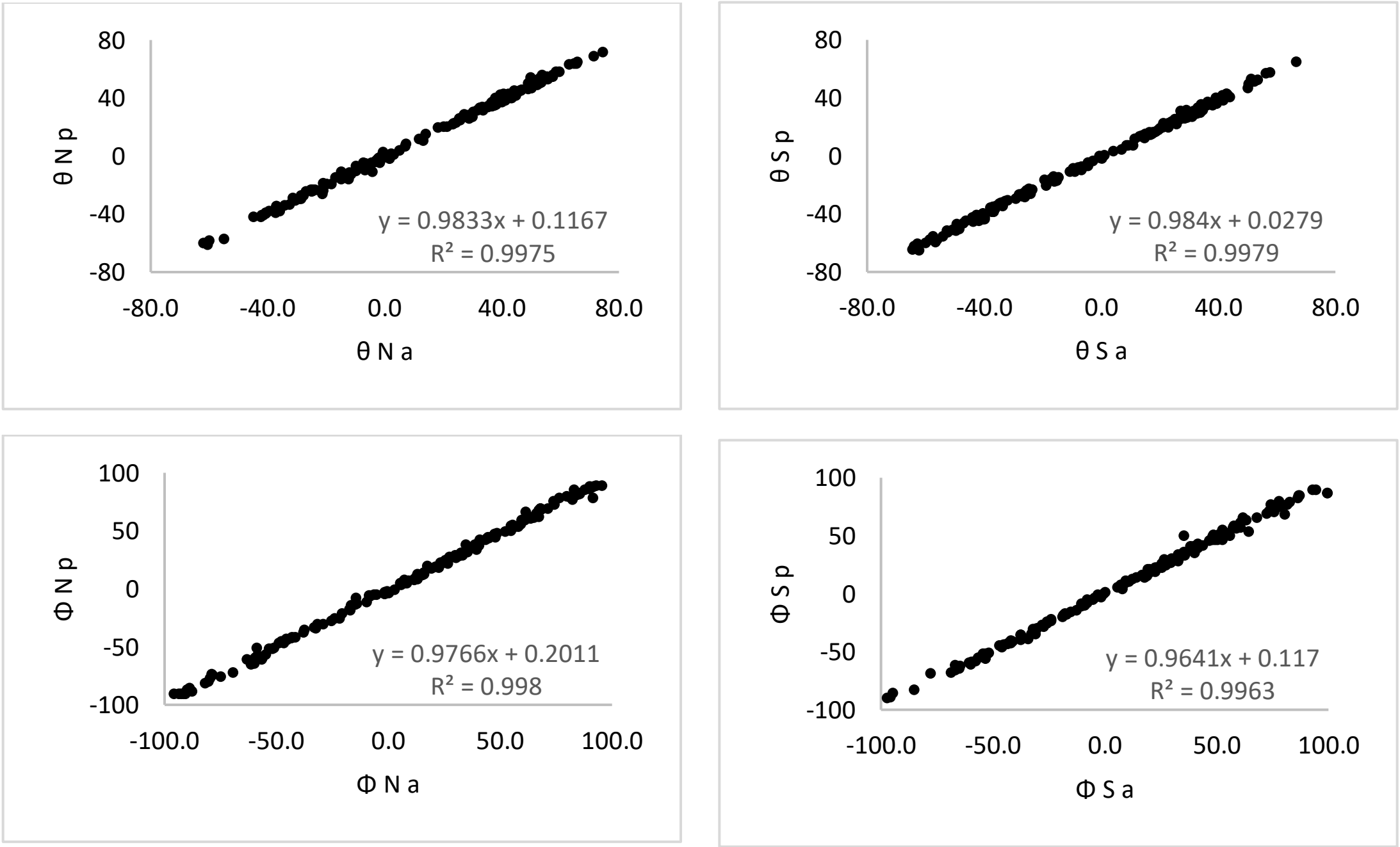

**Figure 9** Linear relationship between coordinates of South and North points published (p) and obtained automatically (a) (θN-latitude of northern points; θS-latitude of southern points; ΦN-longitude of northern points; ΦS-longitude of southern points).

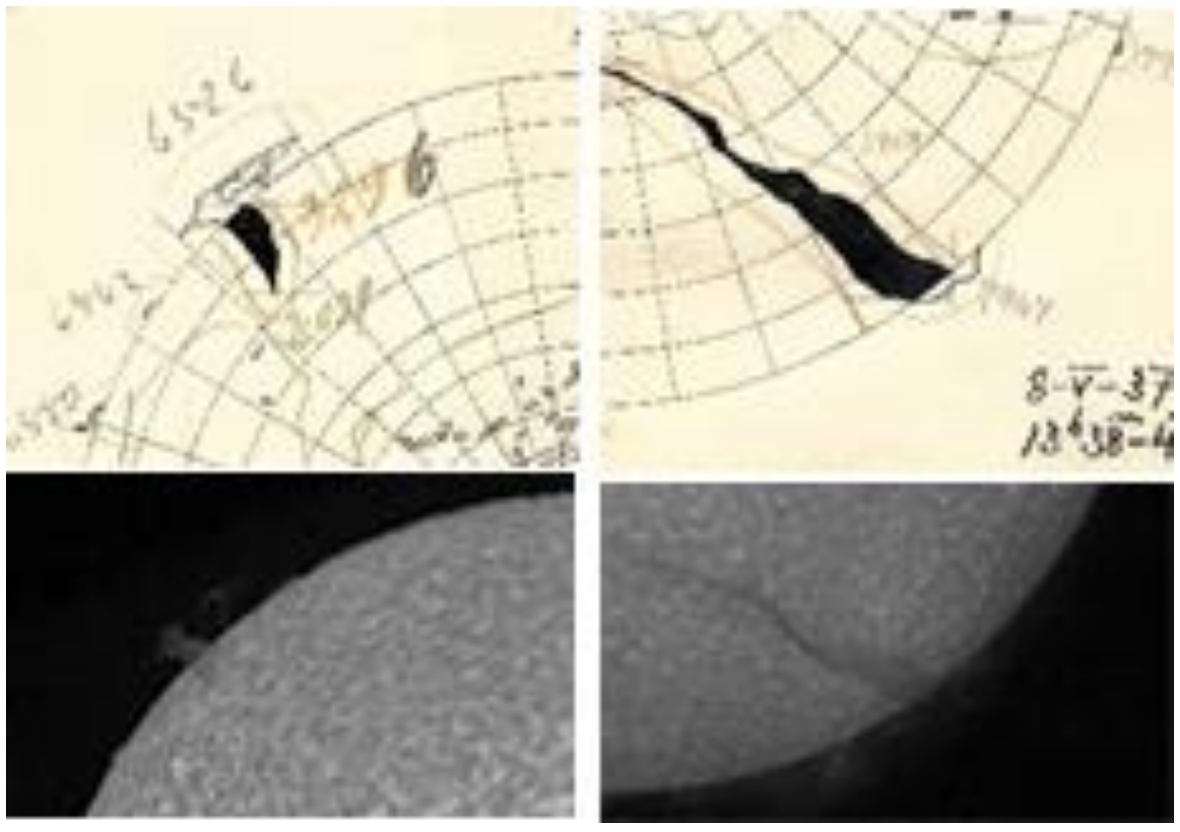

A B

**Figure 10** Drawings and spectroheliograms of 16 September 1935 (A) and of 8 May 1937 (B), where the longitude was calculated for the extension of the filament outside the solar disk.

**Table 1** Basic descriptive statistics of differences observed between heliographic coordinates published and obtained automatically, and between solar hemispheres.

| | | | | | **N** Hemisphere | | **S** Hemisphere | |
|---|---|---|---|---|---|---|---|---|
| | θ N | θ S | Φ N | Φ S | θ N>0 | θ S>0 | θ N<0 | θ S<0 |
| Md | 1.4 | 1.4 | 2.1 | 2.3 | 1.3 | 1.5 | 1.5 | 1.4 |
| Min | 0.0 | 0.0 | 0.0 | 0.0 | 0.0 | 0.0 | 0.0 | 0.0 |
| Max | 6.4 | 7.6 | 11.7 | 14.9 | 4.5 | 5.8 | 6.4 | 5.9 |
| Sd | 1.1 | 1.0 | 1.8 | 2.9 | 0.9 | 1.0 | 1.1 | 1.1 |

*θ N, θ S-northern and southern latitude limits; Φ N and Φ S-northern and southern longitude limits; Md-arithmetic mean; Max-maximum; Min-minimum; Sd-standard deviation.

The comparison with other sources would be another desirable and valuable way to validate the dataset analysed in this paper, but so far, a limited number of historical catalogues is available. On the other hand, the digitized catalogues available mainly contain data on sunspots (*e.g.*, Aparicio et al. 2018; Carrasco et al., 2018b; Carrasco et al., 2014).

## 5. Conclusions

This work aims to validate the results published from the daily CaII K3 observations acquired at the OGAUC, in the early 20$^{th}$ century. At that time, a device developed by an astronomer of the Coimbra astronomical observatory was used to speed up the determination process of heliographic coordinates.

We analysed data on solar filaments coordinates published in the solar catalogues from 1929 to 1941 and compared it with data obtained by an automatic method. The relationship between computed coordinates and coordinates published show a positive and highly significant correlation ($r$=0.99). This high correlation demonstrates that the methodology used at the OGAUC in the early 20$^{th}$ century produced reliable data. The averages of differences between latitudes and longitudes computed and published were calculated and the values obtained are small, 1.2$^0$ for latitudes and 2.4$^0$ for longitudes. The reasons for the higher value for the longitudes may be related to the difficulty in determining features located near the limb and to those cases where longitude was calculated considering the filament extended outside the solar disk. The north-south asymmetry was also analysed to verify if the catalogues returned accurate results in both hemispheres. The difference (mean value) observed is not significant (=0.1$^0$) so it can be concluded that the performance of the method was reliable in both hemispheres. The results demonstrate that data on filaments positions published by the Coimbra Astronomical Observatory in the first half of the twentieth century, contain reliable data and therefore can be considered for studies of solar activity. One of our goals as future work is to identify and correct the errors in the OGAUC catalogues and provide a table of necessary corrections. The digitization of the entire set of catalogues must be completed before this task, which will allow their preservation and the creation of a machine-readable version. This version already exists for sunspots data (Carrasco et al., 2018b) but the OGAUC catalogues contain more information on filaments, protuberances and plages, that should be made available to the scientific community. In this work, we focused only on the position of the solar filaments to verify the reliability of OGAUC's used methodology and, consequently, the catalogues data. Nevertheless, the Coimbra catalogues contain more information about other morphological properties of filaments, like length, thickness and shape, and we expect that after all the data is available, it will contribute valuable information to long-term studies of solar filaments. Although we just have access to the endpoints of the filaments, limiting our interpretation of this individual features, these still provided a global estimation of other quantities upper limits.

**Acknowledgement**: This study was partially supported by the Portuguese Government through the Foundation for Science and Technology–FCT, CITEUC Funds (project: ID/Multi/00611/2013), and FEDER–European Regional Development Fund through COMPETE 2020–Operational Programme Competitiveness and Internationalization (project: POCI–01–0145–FEDER–006922). Ana Lourenço´s work was supported by the project ReNATURE–Valuation of Endogenous Natural Resources in the Central Region (CENTRO–25 01–0145–FEDER–000007–BPD16).

**Disclosure of Potential Conflicts of Interest** The authors declare that they have no conflicts of interest.

## Appendix

**Table caption:** D - day; M - month; Y - year; N - filament number; T - time; JD - Julian day; N θp – northern latitude published; N θa – northern latitude calculated automatically; S θp – southern latitude published; S θa - southern latitude calculated automatically; N Φp – northern longitude published; N Φa – northern longitude calculated automatically; SΦp - southern longitude published; S Φa – southern longitude calculated automatically.

| D | M | Y | N | T | JD | N θp | N θa | S θp | S θa | N Φp | N Φa | S Φp | S Φa |
|---|---|---|---|---|---|---|---|---|---|---|---|---|---|
| 1 | 1 | 1929 | 1 | 09h_57 | 2425613 | 42.8 | 40.0 | 30.8 | 27.7 | 87.6 | 86.4 | 73.1 | 71.4 |
| 1 | 1 | 1929 | 2 | 09h_57 | 2425613 | -18.4 | -19.0 | -33.8 | -34.2 | 25.2 | 25.2 | 73.9 | 77.2 |
| 4 | 1 | 1929 | 1 | 09h_50 | 2425616 | 43.6 | 44.0 | 33.2 | 31.0 | 54.5 | 54.9 | 23.0 | 22.9 |
| 4 | 1 | 1929 | 2 | 09h_50 | 2425616 | *-21.5* | -26.0 | -33.9 | -32.0 | -9.1 | -4.9 | 32.7 | 27.9 |
| 16 | 1 | 1929 | 25 | 10h_00 | 2425628 | -24.0 | -23.5 | -36.9 | -37.8 | 32.3 | 31.0 | 86.6 | 84.6 |
| 28 | 1 | 1930 | 602 | 10h_35 | 2426005 | 52.7 | 51.0 | 33.6 | 30.6 | 82.8 | 80.4 | 42.5 | 41.9 |
| 28 | 1 | 1930 | 603 | 10h_35 | 2426005 | 11.5 | 11.8 | -4.6 | -4.0 | 85.4 | 82.5 | 58.7 | 56.5 |
| 28 | 1 | 1930 | 606 | 10h_35 | 2426005 | -7.7 | -5.0 | -24.8 | -22.1 | -2.0 | -3.5 | 17.6 | 14.4 |
| 11 | 2 | 1930 | 613 | 10h_17 | 2426019 | 27.2 | 27.3 | 6.7 | 5.0 | -15.0 | -12.7 | -32.8 | -33.7 |
| 11 | 2 | 1930 | 618 | 10h_17 | 2426019 | -6.8 | -4.9 | -28.2 | -26.0 | 44.2 | 45.4 | 54.6 | 52.0 |
| 11 | 2 | 1930 | 620 | 10h_17 | 2426019 | 29.7 | 26.9 | 16.6 | 15.8 | 76.0 | 79.0 | *64.2* | 53.8 |
| 12 | 2 | 1931 | 1007 | 10h_02 | 2426385 | 28.5 | 26.0 | 9.1 | 8.0 | 29.7 | 29.6 | 18.9 | 21.2 |
| 25 | 6 | 1931 | 1076 | 10h_02 | 2426518 | 12.8 | 10.8 | 0.1 | -1.4 | 38.1 | 36.9 | 29.3 | 29.0 |
| 21 | 7 | 1931 | 1103 | 09h_04 | 2426544 | 2.8 | 1.1 | -17.2 | -15.3 | *34.4* | 39.0 | 46.5 | 46.2 |
| 9 | 7 | 1932 | 2000 | 09h_25 | 2426898 | -10.2 | -7.1 | -26.6 | -25.1 | -53.6 | -51.0 | -60.3 | -61.0 |
| 4 | 9 | 1932 | 2091 | 10h_50 | 2426955 | 23.0 | 22.0 | 21.0 | 22.8 | **-91.6** | -89.7 | -55.7 | -54.7 |
| 26 | 4 | 1933 | 2366 | 10h_09 | 2427189 | 24.2 | 23.0 | 8.4 | 7.8 | 55.2 | 55.5 | 34.5 | 34.9 |
| 28 | 5 | 1933 | 2407 | 11h_18 | 2427221 | 37.1 | 35.0 | 13.9 | 14.0 | 32.3 | 31.4 | -34.6 | -38.7 |
| 29 | 10 | 1933 | 2607 | 11h_13 | 2427375 | 53.2 | 51.0 | 34.5 | 32.0 | -45.4 | -42.9 | -26.2 | -23.9 |
| 19 | 7 | 1934 | 2849 | 09h_21 | 2427638 | 55.4 | 52.8 | 49.9 | 47.6 | -79.0 | -73.4 | -37.8 | -35.7 |
| 10 | 8 | 1934 | 2893 | 09h_34 | 2427660 | -45.0 | -42.0 | -53.0 | -51.0 | 55.4 | 53.6 | 81.3 | 76.6 |
| 10 | 8 | 1934 | 2892 | 09h_34 | 2427660 | 36.1 | 34.5 | 33.9 | 33.0 | -80.7 | -79.1 | -59.6 | -58.4 |
| 12 | 8 | 1934 | 2893 | 10h_14 | 2427662 | -42.6 | -42.0 | -64.1 | -62.0 | 24.3 | 23.1 | 78.6 | 76.1 |
| 11 | 9 | 1934 | 2930 | 10h_26 | 2427692 | -34.6 | -34.1 | -62.9 | -60.0 | -16.8 | -13.5 | 75.5 | 70.8 |
| 1 | 10 | 1934 | 2955 | 13h_27 | 2427712 | 37.1 | 36.1 | 10.8 | 8.0 | **95.0** | 89.8 | 12.2 | 12.8 |
| 9 | 10 | 1934 | 2955 | 15h_22 | 2427720 | 30.0 | 30.3 | 23.8 | 24.0 | -82.2 | -80.8 | -43.5 | -43.2 |
| 10 | 10 | 1934 | 2955 | 10h_31 | 2427721 | 29.3 | 28.0 | 26.6 | 26.0 | **-93.0** | -90.0 | -54.7 | -51.4 |
| 9 | 3 | 1935 | 3094 | 10h_25 | 2427871 | -40.8 | -39.5 | -49.5 | -46.7 | 5.4 | 4.4 | -32.3 | -30.4 |
| 25 | 3 | 1935 | 3117 | 11h_09 | 2427887 | -41.0 | -39.2 | -50.9 | -50.2 | 51.8 | 49.9 | 67.9 | 65.4 |
| 15 | 4 | 1935 | 3140 | 08h_50 | 2427908 | -13.8 | -13.7 | -34.9 | -32.3 | -60.4 | -61.4 | -37.2 | -37.9 |
| 23 | 4 | 1935 | 3148 | 09h_22 | 2427916 | -31.6 | -28.9 | -43.9 | -45.0 | 22.3 | 20.0 | 72.3 | 68.9 |
| 5 | 7 | 1935 | 3212 | 09h_21 | 2427989 | -29.4 | -29.3 | -32.2 | -30.0 | -75.4 | -75.3 | -45.1 | -43.7 |
| 5 | 7 | 1935 | 3213 | 09h_21 | 2427989 | 25.1 | 24.8 | 21.6 | 21.0 | **-91.0** | -89.7 | -57.9 | -57.7 |
| 5 | 7 | 1935 | 3219 | 09h_21 | 2427989 | -28.0 | -27.3 | -44.1 | -42.6 | 33.8 | 32.3 | 60.5 | 57.3 |
| 24 | 7 | 1935 | 3231 | 11h_50 | 2428008 | -14.1 | -13.3 | -36.4 | -35.6 | -33.3 | -32.3 | -10.0 | -9.5 |
| 5 | 9 | 1935 | 3291 | 09h_51 | 2428051 | -41.8 | -41.0 | -57.8 | -55.0 | 23.0 | 23.5 | 41.4 | 39.7 |
| 5 | 9 | 1935 | 3295 | 09h_51 | 2428051 | -22.5 | -24.0 | -46.5 | -44.0 | 40.0 | 40.5 | 82.6 | 78.8 |
| 5 | 9 | 1935 | 3296 | 09h_51 | 2428051 | 44.0 | 45.0 | 41.3 | 42.0 | **92.5** | 89.5 | 60.3 | 61.5 |
| 7 | 9 | 1935 | 3291 | 09h_47 | 2428053 | -42.2 | -41.0 | -58.8 | -57.0 | -0.9 | -1.8 | 19.8 | 18.5 |
| 7 | 9 | 1935 | 3295 | 09h_47 | 2428053 | -21.0 | -19.7 | -42.9 | -40.0 | 14.3 | 12.7 | 39.9 | 35.1 |
| 12 | 9 | 1935 | 3296 | 09h_00 | 2428058 | 49.4 | 49.1 | 43.0 | 42.7 | -31.9 | -30.0 | -15.5 | -15.5 |
| 12 | 9 | 1935 | 3303 | 09h_00 | 2428058 | 48.7 | 47.0 | 39.2 | 36.7 | 85.0 | 83.0 | 57.0 | 57.0 |
| 16 | 9 | 1935 | 3296 | 10h_18 | 2428062 | 48.7 | 46.2 | 43.7 | 41.2 | **-96.2** | -89.8 | *-78.0* | -68.8 |
| 22 | 9 | 1935 | 3313 | 09h_37 | 2428068 | 52.6 | 53.4 | 42.7 | 43.5 | *-79.2* | -72.9 | -37.9 | -35.1 |
| 29 | 4 | 1936 | 3407 | 10h_23 | 2428288 | -55.1 | -57.4 | -57.0 | -57.0 | 67.0 | 62.6 | -67.0 | -66.0 |
| 29 | 4 | 1936 | 3420 | 10h_23 | 2428288 | -2.1 | -4.5 | -16.4 | -13.6 | 44.7 | 45.1 | 32.0 | 31.2 |
| 29 | 4 | 1936 | 3424 | 10h_23 | 2428288 | 7.0 | 8.2 | -7.0 | -7.0 | -69.9 | -71.6 | -64.9 | -62.0 |
| 29 | 4 | 1936 | 3425 | 10h_23 | 2428288 | 26.5 | 27.8 | 13.9 | 14.3 | -57.0 | -60.0 | -61.0 | -59.5 |
| 30 | 4 | 1936 | 3407 | 10h_26 | 2428289 | -62.1 | -60.0 | -62.5 | -64.5 | 57.6 | 54.4 | **-94.9** | -85.2 |
| 2 | 5 | 1936 | 3407 | 09h_16 | 2428291 | -60.8 | -61.0 | -64.7 | -64.0 | **91.0** | 79.3 | **-97.4** | -89.7 |
| 2 | 5 | 1936 | 3411 | 09h_16 | 2428291 | -15.3 | -16.0 | -39.9 | -41.3 | -52.2 | -51.3 | -28.3 | -27.2 |
| 2 | 5 | 1936 | 3424 | 09h_16 | 2428291 | 2.0 | 1.8 | -4.6 | -6.4 | **-90.2** | -86.5 | -68.8 | -68.2 |
| 2 | 5 | 1936 | 3431 | 09h_16 | 2428291 | 65.4 | 65.1 | 51.0 | 53.6 | -24.4 | -24.7 | -1.8 | -2.8 |
| 2 | 5 | 1936 | 3419 | 09h_16 | 2428291 | -30.8 | -30.6 | -37.8 | -36.6 | -2.1 | -2.4 | 39.4 | 40.9 |
| 2 | 5 | 1936 | 3415 | 09h_16 | 2428291 | 43.3 | 41.5 | 27.4 | 26.5 | 12.6 | 9.3 | -24.4 | -23.7 |
| 9 | 5 | 1936 | 3407 | 14h_32 | 2428298 | -60.2 | -58.5 | -62.9 | -62.0 | -1.8 | -3.5 | **-95.9** | -89.2 |
| 9 | 5 | 1936 | 3438 | 14h_32 | 2428298 | 58.3 | 58.1 | 38.0 | 36.0 | 71.0 | 69.8 | -42.3 | -42.2 |
| 1 | 6 | 1936 | 3474 | 09h_25 | 2428321 | 44.5 | 42.0 | 32.7 | 34.0 | 89.8 | 86.5 | 20.0 | 20.5 |
| 4 | 6 | 1936 | 3474 | 09h_23 | 2428324 | 49.2 | 50.4 | 37.9 | 35.2 | 61.4 | 62.0 | -13.0 | -14.1 |
| 4 | 6 | 1936 | 3476 | 09h_23 | 2428324 | 32.6 | 33.0 | 20.0 | 19.5 | 15.2 | 14.1 | -5.6 | -5.3 |
| 7 | 6 | 1936 | 3491 | 09h_05 | 2428327 | 57.0 | 56.0 | 53.2 | 53.0 | 89.9 | 88.1 | 49.1 | 46.3 |
| 11 | 6 | 1936 | 3491 | 09h_57 | 2428331 | 62.7 | 63.0 | 50.2 | 50.3 | 58.8 | 56.3 | -10.6 | -8.7 |
| 11 | 6 | 1936 | 3446 | 09h_57 | 2428331 | -12.4 | -11.6 | -24.4 | -25.7 | 47.7 | 45.0 | 77.5 | 79.9 |
| 30 | 7 | 1936 | 3567 | 09h_17 | 2428380 | -8.0 | -7.2 | -33.8 | -31.7 | -8.4 | -5.6 | 7.8 | 4.2 |
| 1 | 8 | 1936 | 3567 | 08h_52 | 2428382 | 4.6 | 4.0 | -27.9 | -25.9 | -47.0 | -46.2 | -26.0 | -25.8 |
| 1 | 8 | 1936 | 3576 | 08h_52 | 2428382 | 65.3 | 63.7 | 57.6 | 58.0 | -61.8 | -64.7 | 18.0 | 14.9 |
| 1 | 8 | 1936 | 3577 | 08h_52 | 2428382 | 37.7 | 35.6 | 26.3 | 26.0 | -51.4 | -50.2 | -57.0 | -55.1 |
| 20 | 8 | 1936 | 3602 | 10h_00 | 2428401 | 6.8 | 7.0 | -0.7 | -0.7 | -43.5 | -41.4 | -18.7 | -18.5 |
| 22 | 8 | 1936 | 3610 | 10h_12 | 2428403 | -4.7 | -4.7 | -23.9 | -23.0 | 0.0 | -2.8 | 19.4 | 20.4 |
| 1 | 9 | 1936 | 3638 | 14h_00 | 2428413 | *-15.2* | -11.0 | -26.4 | -27.9 | 27.0 | 27.9 | 52.4 | 55.1 |
| 15 | 9 | 1936 | 3650 | 09h_44 | 2428427 | 57.0 | 54.9 | 56.2 | 57.7 | 47.0 | 47.7 | -31.0 | -32.2 |
| 15 | 9 | 1936 | 3666 | 09h_44 | 2428427 | 40.6 | 42.2 | 27.1 | 30.0 | 67.7 | 70.0 | 48.5 | 50.6 |
| 2 | 10 | 1936 | 3691 | 09h_21 | 2428444 | -21.3 | -18.7 | -34.3 | -32.0 | 55.4 | 54.3 | 25.2 | 25.9 |
| 2 | 10 | 1936 | 3694 | 09h_21 | 2428444 | 51.4 | 51.0 | 38.7 | 37.9 | -14.4 | -12.1 | -24.2 | -21.9 |
| 7 | 10 | 1936 | 3695 | 10h_07 | 2428449 | 41.8 | 39.8 | 31.5 | 31.6 | 40.2 | 37.8 | 13.9 | 14.4 |
| 7 | 10 | 1936 | 3700 | 10h_07 | 2428449 | 33.0 | 32.0 | 24.9 | 25.9 | 65.8 | 65.9 | 52.9 | 53.5 |
| 7 | 10 | 1936 | 3701 | 10h_07 | 2428449 | 55.1 | 54.5 | 52.2 | 52.0 | 61.0 | 67.0 | *35.1* | 50.0 |
| 16 | 10 | 1936 | 3704 | 10h_52 | 2428458 | 71.1 | 69.0 | 17.1 | 15.7 | 26.4 | 22.6 | -4.7 | -3.7 |
| 16 | 10 | 1936 | 3714 | 10h_52 | 2428458 | 41.3 | 41.0 | 26.5 | 26.0 | 80.8 | 80.0 | **94.0** | 89.6 |
| 21 | 10 | 1936 | 3717 | 10h_14 | 2428463 | -20.1 | -19.4 | -40.5 | -39.3 | -17.6 | -16.9 | -10.7 | -10.4 |
| 21 | 10 | 1936 | 3726 | 10h_14 | 2428463 | -21.4 | -23.8 | -44.2 | -41.9 | 43.0 | 42.9 | *80.3* | 68.2 |
| 21 | 10 | 1936 | 3727 | 10h_14 | 2428463 | 43.2 | 40.0 | 27.1 | 26.4 | 11.2 | 8.1 | 0.0 | 1.6 |
| 5 | 12 | 1936 | 3794 | 10h_34 | 2428508 | 49.6 | 50.4 | 28.7 | 31.1 | 14.8 | 12.7 | -9.1 | -9.8 |
| 6 | 1 | 1937 | 3838 | 11h_33 | 2428540 | 32.3 | 33.4 | 12.9 | 13.7 | 60.8 | 60.2 | 29.5 | 30.3 |
| 7 | 1 | 1937 | 3838 | 11h_04 | 2428541 | 33.0 | 33.6 | 11.3 | 12.3 | 40.4 | 42.7 | 10.5 | 10.4 |
| 20 | 2 | 1937 | 3878 | 15h_00 | 2428585 | 52.7 | 50.0 | 30.4 | 28.4 | *-14.8* | -7.2 | 18.8 | 15.4 |
| 25 | 3 | 1937 | 3918 | 10h_03 | 2428618 | -29.0 | -29.2 | -56.9 | -59.0 | 35.2 | 32.4 | **99.4** | 86.5 |
| 25 | 3 | 1937 | 3921 | 10h_03 | 2428618 | -7.6 | -4.6 | *-19.6* | -16.0 | -32.8 | -33.1 | -27.6 | -28.1 |
| 25 | 3 | 1937 | 3922 | 10h_03 | 2428618 | *49.5* | 54.0 | 34.0 | 36.1 | *-59.3* | -50.5 | -52.0 | -51.0 |
| 26 | 3 | 1937 | 3913 | 10h_38 | 2428619 | 41.0 | 38.1 | 26.9 | 31.3 | 31.2 | 28.7 | 9.2 | 11.2 |
| 26 | 3 | 1937 | 3918 | 10h_38 | 2428619 | -28.8 | -28.3 | -56.3 | -57.4 | 20.7 | 19.8 | **99.2** | 86.8 |
| 22 | 4 | 1937 | 3941 | 10h_05 | 2428646 | -39.9 | -38.0 | -41.4 | -40.1 | 47.0 | 45.9 | 74.5 | 73.9 |
| 22 | 4 | 1937 | 3942 | 10h_05 | 2428646 | -3.2 | -2.8 | -9.1 | -10.5 | 39.3 | 38.7 | 47.8 | 49.2 |
| 23 | 4 | 1937 | 3941 | 10h_12 | 2428647 | -37.3 | -37.4 | -42.6 | -42.1 | 30.1 | 27.5 | 49.9 | 50.1 |
| 23 | 4 | 1937 | 3942 | 10h_12 | 2428647 | -1.5 | -0.3 | -16.2 | -16.9 | 23.5 | 23.4 | 36.0 | 35.3 |
| 26 | 4 | 1937 | 3949 | 09h_53 | 2428650 | 39.4 | 38.1 | 29.3 | 26.9 | 64.9 | 62.2 | 26.7 | 27.6 |
| 28 | 4 | 1937 | 3955 | 09h_19 | 2428652 | -20.3 | -19.5 | -37.4 | -37.9 | 37.3 | 36.8 | 61.6 | 65.4 |
| 1 | 5 | 1937 | 3941 | 08h_34 | 2428655 | -0.8 | -1.0 | -40.0 | -43.0 | *-89.1* | -85.0 | -53.3 | -53.0 |
| 1 | 5 | 1937 | 3956 | 08h_34 | 2428655 | 20.0 | 20.3 | 1.0 | 0.9 | -43.7 | -41.7 | -31.1 | -30.6 |
| 1 | 5 | 1937 | 3963 | 08h_34 | 2428655 | -1.9 | -2.3 | -6.9 | -9.2 | 29.1 | 28.4 | 16.4 | 15.2 |
| 1 | 5 | 1937 | 3965 | 08h_34 | 2428655 | 25.1 | 25.4 | 16.2 | 16.9 | -49.1 | -46.0 | -66.3 | -63.0 |
| 2 | 5 | 1937 | 3949 | 10h_26 | 2428656 | 46.5 | 45.6 | 31.5 | 31.0 | -6.9 | -4.2 | -53.4 | -51.3 |
| 2 | 5 | 1937 | 3956 | 10h_26 | 2428656 | 17.9 | 19.9 | -0.7 | 0.4 | -55.2 | -55.9 | -46.3 | -45.7 |
| 8 | 5 | 1937 | 3968 | 13h_38 | 2428662 | -37.2 | -34.3 | -54.4 | -54.8 | 2.6 | 0.1 | **92.8** | 89.5 |
| 23 | 5 | 1937 | 3982 | 10h_32 | 2428677 | -32.9 | -33.2 | -43.7 | -43.2 | 6.7 | 8.1 | 50.0 | 46.4 |
| 12 | 6 | 1937 | 4006 | 10h_06 | 2428697 | -37.7 | -39.0 | -40.6 | -43.1 | -88.1 | -87.6 | -41.3 | -40.7 |
| 29 | 6 | 1937 | 4032 | 13h_10 | 2428714 | -28.8 | -27.0 | -43.9 | -42.0 | 39.9 | 39.4 | 86.7 | 84.4 |
| 1 | 7 | 1937 | 4031 | 09h_38 | 2428716 | 42.0 | 43.1 | 14.9 | 15.7 | 59.2 | 59.9 | 19.8 | 20.4 |
| 1 | 7 | 1937 | 4032 | 09h_38 | 2428716 | -25.2 | -24.2 | -42.9 | -42.0 | 15.5 | 14.9 | 86.0 | 82.9 |
| 18 | 7 | 1937 | 4055 | 09h_25 | 2428733 | 1.4 | -0.3 | -6.2 | -7.1 | -29.5 | -29.7 | -47.2 | -44.4 |
| 19 | 7 | 1937 | 4055 | 09h_45 | 2428734 | 0.0 | 0.9 | -8.2 | -7.4 | -45.8 | -42.3 | -65.3 | -64.3 |
| 20 | 7 | 1937 | 4066 | 10h_21 | 2428735 | -11.8 | -12.5 | -27.2 | -27.2 | 12.7 | 13.4 | 26.2 | 25.2 |
| 2 | 9 | 1937 | 4130 | 09h_32 | 2428779 | 74.2 | 72.0 | 66.3 | 65.6 | -63.6 | -60.4 | -8.2 | -4.8 |
| 3 | 10 | 1937 | 4203 | 10h_25 | 2428810 | 53.6 | 55.9 | 39.0 | 40.5 | 48.4 | 48.5 | 19.6 | 21.2 |
| 29 | 11 | 1937 | 4241 | 11h_00 | 2428867 | 37.1 | 37.0 | 27.0 | 27.4 | 83.0 | 82.9 | 50.5 | 51.1 |
| 29 | 11 | 1937 | 4242 | 11h_00 | 2428867 | 48.7 | 50.0 | 41.3 | 39.3 | 73.6 | 75.9 | 43.5 | 41.5 |
| 22 | 12 | 1937 | 4156 | 10h_32 | 2428890 | 25.1 | 25.7 | 17.5 | 16.4 | 65.0 | 63.8 | 41.5 | 42.7 |
| 22 | 12 | 1937 | 4157 | 10h_32 | 2428890 | 50.6 | 49.7 | 40.2 | 39.0 | 67.6 | 69.6 | 52.4 | 46.8 |
| 22 | 12 | 1937 | 4159 | 10h_32 | 2428890 | 39.1 | 42.2 | 36.1 | 37.9 | 43.4 | 43.8 | 28.3 | 27.8 |
| 4 | 1 | 1938 | 4168 | 10h_07 | 2428903 | 35.0 | 34.0 | 25.4 | 22.4 | 39.0 | 34.6 | 25.3 | 22.8 |
| 6 | 2 | 1938 | 4185 | 11h_26 | 2428936 | 49.7 | 47.0 | 32.3 | 30.0 | 79.5 | 80.1 | 52.3 | 49.7 |
| 8 | 2 | 1938 | 4191 | 10h_19 | 2428938 | 59.4 | 58.0 | 27.8 | 27.0 | 82.0 | 77.8 | 22.5 | 22.8 |
| 10 | 2 | 1938 | 4191 | 09h_57 | 2428940 | 64.3 | 63.6 | 33.2 | 31.0 | 54.5 | 50.9 | -3.5 | -1.0 |
| 2 | 3 | 1938 | 4208 | 11h_22 | 2428960 | -3.8 | -4.1 | -19.0 | -20.0 | 47.2 | 46.2 | 27.0 | 24.6 |
| 2 | 3 | 1938 | 4209 | 11h_22 | 2428960 | 33.3 | 31.8 | 3.9 | 4.0 | 44.5 | 44.7 | 7.0 | 7.9 |
| 9 | 3 | 1938 | 4218 | 10h_01 | 2428967 | -14.1 | -14.5 | -41.0 | -41.3 | 22.2 | 20.2 | 74.2 | 73.3 |
| 9 | 4 | 1938 | 4296 | 09h_48 | 2428998 | -12.6 | -16.0 | -15.5 | -16.7 | 32.7 | 29.7 | 58.1 | 56.9 |
| 11 | 6 | 1938 | 4387 | 09h_23 | 2429061 | -1.0 | 2.8 | -14.7 | -14.3 | -17.3 | -17.8 | 6.8 | 6.1 |
| 20 | 6 | 1938 | 4423 | 10h_00 | 2429070 | 39.6 | 41.0 | 16.6 | 15.0 | **91.5** | 88.9 | 57.6 | 58.4 |
| 13 | 7 | 1938 | 4453 | 10h_37 | 2429093 | 46.1 | 45.4 | 41.8 | 41.3 | 82.6 | 86.0 | 29.9 | 28.1 |
| 13 | 7 | 1938 | 4455 | 10h_37 | 2429093 | 40.2 | 39.6 | 35.9 | 35.7 | -79.9 | -76.1 | -45.4 | -43.6 |
| 14 | 7 | 1938 | 4455 | 14h_24 | 2429094 | 38.9 | 38.4 | 38.0 | 36.7 | **-93.8** | -89.8 | -67.2 | -61.7 |
| 26 | 8 | 1938 | 4527 | 09h_53 | 2429137 | 38.9 | 39.0 | 17.3 | 16.9 | 67.1 | 68.7 | 33.5 | 31.6 |
| 30 | 8 | 1938 | 4528 | 09h_41 | 2429141 | -28.6 | -28.0 | -60.1 | -59.3 | -25.9 | -27.0 | 77.3 | 74.9 |
| 30 | 8 | 1938 | 4536 | 09h_41 | 2429141 | 1.4 | -1.6 | -15.7 | -15.0 | 19.1 | 18.3 | 16.6 | 16.4 |
| 16 | 10 | 1938 | 4588 | 10h_43 | 2429188 | -2.6 | -1.8 | -9.7 | -7.8 | 38.2 | 39.0 | 47.8 | 46.3 |
| 16 | 10 | 1938 | 4589 | 10h_43 | 2429188 | 35.0 | 34.3 | 19.5 | 18.6 | 7.3 | 5.4 | -33.0 | -34.9 |
| 16 | 10 | 1938 | 4590 | 10h_43 | 2429188 | -24.6 | -24.0 | -29.5 | -29.0 | 15.8 | 13.3 | 32.7 | 33.9 |
| 7 | 11 | 1938 | 4625 | 10h_15 | 2429210 | -6.9 | -9.8 | -11.0 | -10.3 | 5.0 | 5.7 | -17.1 | -17.8 |
| 7 | 11 | 1938 | 4629 | 10h_15 | 2429210 | 23.3 | 22.7 | 21.9 | 21.3 | 61.5 | 62.4 | 38.1 | 40.9 |
| 13 | 11 | 1938 | 4629 | 14h_29 | 2429216 | 25.4 | 24.9 | 20.7 | 20.1 | -42.0 | -40.9 | -30.3 | -29.4 |
| 13 | 11 | 1938 | 4631 | 14h_29 | 2429216 | 31.4 | 31.8 | 21.2 | 21.4 | -5.7 | -4.4 | -25.2 | -24.3 |
| 13 | 11 | 1938 | 4632 | 14h_29 | 2429216 | 37.6 | 36.9 | 30.1 | 29.2 | -21.0 | -20.6 | -42.0 | -40.2 |
| 13 | 11 | 1938 | 4635 | 14h_29 | 2429216 | -17.2 | -14.7 | -25.7 | -23.5 | 38.8 | 35.9 | 63.1 | 63.6 |
| 13 | 11 | 1938 | 4640 | 14h_29 | 2429216 | 45.8 | 45.4 | 41.4 | 39.0 | 8.8 | 7.5 | -8.0 | -7.8 |
| 2 | 2 | 1939 | 4700 | 10h_24 | 2429297 | 44.7 | 43.5 | 36.3 | 37.0 | 84.2 | 82.1 | 35.2 | 36.2 |
| 2 | 2 | 1939 | 4704 | 10h_24 | 2429297 | -36.1 | -37.9 | -42.0 | -44.0 | -60.2 | -63.9 | -43.2 | -42.6 |
| 2 | 3 | 1939 | 4733 | 10h_39 | 2429325 | 40.4 | 43.0 | 38.1 | 36.6 | 68.1 | 69.4 | 52.9 | 49.3 |
| 2 | 3 | 1939 | 4735 | 10h_39 | 2429325 | -27.2 | -24.4 | -47.3 | -45.7 | 11.1 | 9.0 | 75.4 | 74.0 |
| 6 | 3 | 1939 | 4735 | 10h_30 | 2429329 | *-4.6* | -11.0 | -49.7 | -51.1 | -48.2 | -44.9 | 16.7 | 15.3 |
| 6 | 3 | 1939 | 4738 | 10h_30 | 2429329 | 51.9 | 49.0 | 29.0 | 31.9 | 61.6 | 61.2 | -37.7 | -39.7 |
| 12 | 5 | 1939 | 4808 | 08h_44 | 2429396 | 36.3 | 34.2 | 22.7 | 20.4 | 73.8 | 73.1 | 35.6 | 33.4 |
| 27 | 6 | 1939 | 4843 | 08h_48 | 2429442 | 13.8 | 15.0 | -3.0 | -3.0 | -58.8 | -56.7 | -27.9 | -27.1 |
| 6 | 7 | 1939 | 4861 | 09h_05 | 2429451 | -37.3 | -34.7 | -48.8 | -50.0 | -22.2 | -24.9 | 8.9 | 8.9 |
| 8 | 7 | 1939 | 4861 | 10h_03 | 2429453 | -25.5 | -23.0 | -52.5 | -52.0 | -56.4 | -59.0 | -19.1 | -19.6 |
| 5 | 8 | 1939 | 4897 | 08h_30 | 2429481 | -10.2 | -10.0 | -37.2 | -34.4 | 17.2 | 20.3 | 29.3 | 26.7 |
| 14 | 9 | 1939 | 4964 | 10h_14 | 2429521 | 40.2 | 40.3 | 30.2 | 30.6 | 22.1 | 19.2 | 5.5 | 5.3 |
| 12 | 10 | 1939 | 4982 | 09h_26 | 2429549 | 36.5 | 37.0 | 26.1 | 26.0 | 89.8 | 88.6 | 51.8 | 51.7 |
| 5 | 4 | 1941 | 5701 | 09h_48 | 2430090 | 37.5 | 36.5 | 28.2 | 26.4 | 63.1 | 61.3 | 50.5 | 47.9 |
| 11 | 4 | 1941 | 5708 | 08h_53 | 2430096 | 39.7 | 37.1 | 32.6 | 29.6 | 85.0 | 82.6 | *55.5* | 50.0 |
| 11 | 4 | 1941 | 5710 | 08h_53 | 2430096 | 28.2 | 28.1 | 18.1 | 17.0 | -10.0 | -10.4 | -18.0 | -17.3 |
| 18 | 4 | 1941 | 5713 | 09h_23 | 2430103 | -24.9 | -23.0 | -32.9 | -30.8 | 12.3 | 11.3 | -0.9 | 0.0 |
| 14 | 6 | 1941 | 5764 | 08h_58 | 2430160 | 36.0 | 36.6 | 18.4 | 17.4 | **90.8** | 88.1 | 22.2 | 18.9 |
| 7 | 8 | 1941 | 5847 | 08h_50 | 2430214 | 27.0 | 28.8 | 22.8 | 23.2 | -38.2 | -36.9 | -53.4 | -55.6 |
| 7 | 8 | 1941 | 5849 | 08h_50 | 2430214 | 37.5 | 39.8 | 22.8 | 22.6 | 7.8 | 5.6 | 26.5 | 29.7 |
| 18 | 8 | 1941 | 5857 | 10h_09 | 2430225 | 21.1 | 20.2 | 14.5 | 12.6 | -59.9 | -58.6 | -85.3 | -82.7 |
| 8 | 12 | 1941 | 5988 | 10h_36 | 2430337 | -2.4 | -1.0 | -36.1 | -34.0 | -38.0 | -34.8 | 6.8 | 6.0 |
| 11 | 12 | 1941 | 5988 | 10h_33 | 2430340 | -6.2 | -7.0 | -38.1 | -35.0 | -89.8 | -87.0 | -31.2 | -34.6 |